\documentstyle[epsfig,preprint,aps]{revtex}     
\begin{document}     
\tighten
\def\bea{\begin{eqnarray}}
\def\eea{\end{eqnarray}} 
\def\beas{\begin{eqnarray*}} 
\def\eeas{\end{eqnarray*}}
\def\nn{\nonumber}
\def\ni{\noindent}
\def\G{\Gamma}
\def\L{\Lambda}
\def\d{\delta}
\def\l{\lambda}
\def\g{\gamma}
\def\m{\mu}
\def\n{\nu}
\def\w{\omega}             
\def\s{\sigma}          
\def\tt{\theta}          
\def\b{\beta}          
\def\a{\alpha}          
\def\f{\phi}          
\def\fh{\phi}          
\def\y{\psi}          

\def\z{\zeta}          
\def\p{\pi}     
\def\h{\hbar}     
\def\e{\epsilon}          
\def\ve{\varepsilon}
\def\cf{{\cal F}}  
\def\cg{{\cal G}}  
\def\ch{{\cal H}}
\def\ci{{\cal I}}            
\def\cl{{\cal L}}         
\def\cv{{\cal V}}         
\def\cz{{\cal Z}}          
\def\pl{\partial}         
\def\ov{\over}          
\def\~{\tilde}         
\def\rar{\rightarrow}          
\def\lar{\leftarrow}          
\def\lrar{\leftrightarrow}          
\def\rra{\longrightarrow}          
\def\lla{\longleftarrow}          
\def\8{\infty}          
\newcommand{\fr}{\frac}

%\twocolumn[\hsize\textwidth\columnwidth\hsize\csname@twocolumnfalse%             
%\endcsname         

\title{A comment on W. Greiner's ``Quantum Mechanics -- An Introduction''}

\author{J. -M. Chung
%\footnote{Email address: jmchung@khu.ac.kr}
}
\address{Research Institute for Basic Sciences
and Department of Physics,\\ Kyung Hee University, Seoul 130-701, Korea}

\maketitle             
\draft            
\begin{abstract}                 
It is pointed out that in Greiner's book, 
``Quantum Mechanics --An
Introduction,''  there is a confusion between 
the wavelength spectral function and the (angular)
frequency spectral function in the derivation of the Wien's
displacement law.
\end{abstract}       

~\\
   
~\\                
%\pacs{PACS number(s): }           
%]          

In his book, Quantum Mechanics -- An Introduction \cite{gr}, Greiner
derives Wien's displacement law from  Planck's spectral energy density
distribution.
In his derivation, there is a mistake due to a confusion between 
the wavelength spectral function and the (angular) frequency
spectral function. For convenience of discussion, here we quote the
problem posed and a part of the solution to it.

\begin{quote}
{\bf Problem.} Derive Wien's displacement law, i.e., 
\beas
\l_{\rm max} T = \rm {const.}
\eeas
from  Planck's spectral energy density $ {1 \ov V} dE/d \w$. 
\underline{ Here $\l_{\rm max} $  is the wavelength} 
\underline{where $ {1 \ov V} dE/d\w$ achieves
its maximum.}  Interpret the result.

{\bf Solution.} We are looking for the maximum of the Planck's spetral
distribution:
\bea
{d \ov d\w} \biggl[ {1\ov V}{dE\ov d \w} \biggr]
&=&{d \ov d\w}\biggl[ {\h\w^3\ov \p^2c^3}\biggl(\exp\biggl( {\h\w\ov k_B
T}\biggr)-1\biggr)^{-1} \biggr]\nn\\
&=& {3\h\w^2\ov \p^2 c^3} \biggl[\exp\biggl( {\h\w\ov k_B
T}\biggr)-1\biggr]^{-1}\nn\\
&&-{\h\w^3\ov\p^2 c^3}{\h\ov k_B T} {\exp(\h\w/k_B T)\ov
[\exp(\h\w/k_B T)-1]^2}=0\nn\\
&\Rightarrow &3 - {\h \w\ov k_B T}\exp\biggl({\h\w\ov k_B
T}\biggr) \biggl[\exp\biggl( {\h\w\ov k_B
T}\biggr)-1\biggr]^{-1} =0\,.
\eea
With the shorthand notation $x={\h\w\ov k_B T}$, we get the transcendental
equation
\bea
e^x =\biggl(1-{x \ov 3} \biggr)^{-1}\;,  \label{e3}
\eea
which must be solved graphically or numerically. Besides the trivial
solution $x=0$ (minimum), a positive solution  exists. Therefore
\bea
x_{\rm max}={\h\w_{\rm max}\ov k_B T}\, ,
\eea
and because \fbox{$\w_{\rm max}=2\p \n_{\rm max} =2\p c/\l_{\rm max}$} 
we have 
\bea
\l_{\rm max} T = \rm {const.}=0.29 ~{\rm cm \,K}\,.\label{wien}
\eea
This means $\cdots\cdots$
\end{quote}

~\\

The Planck's spectral distribution used in the above solution 
\bea
{1 \ov V}{d E(\w,T)\ov d\w}= {\h\w^3\ov \p^2c^3}\biggl[\exp\biggl( 
{\h\w\ov k_B T}\biggr)-1\biggr]^{-1} \equiv \tilde{u} (\w, T) \label{ut}
\eea
is the angular frequency spectral function. The wavelength and
frequency spectral
functions take the following forms:
\bea
u(\l, T)&=& {8\p h c\ov \l^5}\biggl[\exp\biggl( {c h\ov k_B
\l T}\biggr)-1\biggr]^{-1}\;,  \label{u}\\
\hat{u}(\n, T)&=& {8\p h \n^3\ov c^3}\biggl[\exp\biggl( { h\n \ov k_B
T}\biggr)-1\biggr]^{-1}\;.\label{uh}\nn
\eea
Physically, these three spectral functions are related by the
following equation:
\beas
u(\l, T) d\l = -\hat{u}(\n,T)d\n =
-\tilde{u}(\w, T) d\w \;.
\eeas

Let us define $\l_{\rm M}$, $\n_{\rm M}$, and $\w_{\rm M}$ as the
wavelength, frequency, and angular frequency at which $u(\l,T)$,
$\hat{u}(\n,T)$, and $\tilde{u}(\w,T)$  have their maximum value.
(At this stage, it is very clear that the `$\w_{\rm max}$' in Eq. (3) is
the same as $\w_M$.) For $\l_{\rm M}$, one must solve the following 
transendental equation
\bea
e^x =\biggl(1-{x \ov 5} \biggr)^{-1}\;,  \label{e5}
\eea
with $x={ c h\ov k_B \l T} $. The equation to solve for $\n_{\rm max}$ is
the same as the one in Eq.~(\ref{e3}), this time, with $x= {h\n \ov  k_B T}$.
The solutions to Eqs. (\ref{e3}) and (\ref{e5}) ($x_1$ and $x_2$,
respectively) are given as follows:
\bea
x_1&=&2.281\cdots\;,\label{x1}\\
x_2&=&4.965\cdots\;. \label{x2}
\eea
Eqs. (\ref{x1}) and (\ref{x2}) yield 
\beas
&&{2\p c \,T\ov \w_{\rm M}}={c\,T\ov \n_{\rm M}}=
{ch\ov k_B x_1}=0.510 ~{\rm cm\,K}\;,\\
&&\,\l_{\rm M}T={ch\ov k_B x_2}=0.289 ~{\rm cm\,K}\;. 
\eeas
Thus, we have
\beas
\w_{\rm M}= 2\p \n_{\rm M} = {2.281\ov 4.965}
\times 2\p c/ \l_{\rm M}\;.
\eeas
If the `$\l_{\max}$' in the underlined sentence (in the quoted problem)
is defined as in the boxed equation above Eq. (\ref{wien}), 
in conformity with
the following general relation:
\beas
\w= 2\p \n = 2\p c/ \l\;,
\eeas
then, we have
\bea
\l_{\rm max} T =2.177\,\, \l_M T =0.629 ~{\rm cm \,K}\,,\label{nwien}
\eea
which differs from Eq.~(\ref{wien}).
   
In conclusion, if the author of \cite{gr} insists that the
`$\l_{\max}$' in Eq.~(\ref{wien})
be the wavelength at which the wavelength spectral function $u(\l, T)$ 
achieves its maximum, in order
to keep the number $0.29$ on the right-hand side of Eq.~(\ref{wien}) as
usual \cite{lb}, 
then the wavelength spectral function $u(\l,T)$ of Eq.~(\ref{u}),
instead of the (angular) frequency spectral function,
should be used.


\begin{thebibliography}{99}

\bibitem{gr}
W. Greiner, {\it Quantum Mechanics --- An Introduction}, 4th
ed. (Springer, Berlin, 2001), p. 24.
 
\bibitem{lb}
For example, R. L. Liboff, {\it Introductory Quantum Mechanics}, 3rd. ed. 
(Addison-Wesley, 1998), Problem 2.4 \,on  p. 35.

\end{thebibliography}
\end{document}